\documentclass[12pt,preprint]{aastex} 
\usepackage{graphicx} 
\usepackage{epstopdf}
\usepackage{natbib}
\usepackage{longtable}

\newcommand{\skipthis}[1]{}

\def\nh3{$\rm{NH_3}$}
\def\NH3{$\rm{NH_3}$} 
 
\def\msun{$M_\odot$}

\def\kms-1{km~s$^{-1}$}
\def\h2{$\rm{H_2}$} 
\def\CM2{$\rm{cm^{-2}}$}
\def\cm3{$\rm{cm^{-3}}$} 

\def\h2co{H$_2$CO} 
\def\hc3n{HC$_3$N} 
\def\vlsr{$V_{\rm{LSR}}$}
\def\n2hp{N$_2$H$^+$}
\def\n2dp{N$_2$D$^+$}
\def\ch3oh{CH$_3$OH}
\def\c18o{C$^{18}$O}

\begin{document}


\title{ Angular Momentum in Disk Wind Revealed in the Young Star MWC349A}

\author{Qizhou Zhang\altaffilmark{1}, Brian Claus\altaffilmark{1}, Linda Watson\altaffilmark{2,1}, James Moran\altaffilmark{1}}

\altaffiltext{1}{Harvard-Smithsonian Center for Astrophysics, 60 Garden Street,
Cambridge MA 02138, USA. E-mail: qzhang@cfa.harvard.edu}


\altaffiltext{2}{European Southern Observatory, Chile}


\begin{abstract}
Disk winds are thought to play a critical role in star birth. As winds extract excess angular momentum from accretion disks, matter in the disk can be transported inward to the star to fuel mass growth. However, the observational evidence of wind carrying angular momentum has been very limited. We present Submillimeter Array (SMA) observations of the young star MWC349A in the H26$\alpha$ and H30$\alpha$ recombination lines. The high signal-to-noise ratios made possible by the maser emission process allow us to constrain the relative astrometry of the maser spots to a milli-arcsecond precision. Previous observations of the H30$\alpha$ line with the SMA and the Plateau de Bure interferometer (PdBI) showed that masers are distributed in the disk and wind.  Our new high resolution observations of the H26$\alpha$ line reveal differences in spatial distribution from that of the H30$\alpha$ line.
H26$\alpha$ line masers in the disk are excited in a thin annulus with a radius of  about 25 AU,  while  the H30$\alpha$ line masers are formed in a slightly larger annulus with a radius of  30 AU. This is consistent with expectations for maser excitation in the presence of an electron density variation of approximately $R^{-4}$. In addition, the H30$\alpha$ and H26$\alpha$ line masers arise from different parts in the wind. This difference is also expected from the maser theory. The wind component of both masers exhibits line-of-sight velocities that closely follow a Keplerian law.  This result provides strong evidence that the disk wind extracts significant angular momentum and thereby facilitating mass accretion in the young star.
\end{abstract}

\section{Introduction}

The formation of a flattened disk surrounding a young stellar object is a natural outcome of a rotating and collapsing molecular core.  A  core with initial angular momentum  leads to a flattened structure as material is accreted toward the  protostar while conserving angular momentum. One of the puzzles in star formation is the so called ``angular momentum problem''. As matter is accreted toward the star, it has  to shed significant angular momentum in order for the accretion to continue. Disks, observed in protostellar systems of a wide range of masses \citep[e.g.][]{beuther2005c,zhang2005b,cesaroni2007}, may play a pivotal role in the mass growth of protostars by extracting excess angular momentum in the infalling matter through winds and outflows along the polar direction \citep{arce2007,zhang2001,zhang2005a, qiu2008}. 
In theoretical models of star formation, magneto hydrodynamic winds extract angular momentum from the disk \citep{shu2000, konigl2000}. As a consequence, matter can be accreted onto the forming star. While theories on protostellar wind do not agree on the location at which the wind is launched,  the role it plays in shedding angular momentum is a common feature in all theoretical models. 

Despite the theoretical consensus,  observational confirmation of wind angular momentum has been elusive for decades. High spatial resolution differential spectroscopy of optical jets in T Tauri stars showed early promise in revealing transverse velocity gradients indicative of jet rotation \citep[e.g.][]{coffey2004,coffey2008}. However, many of the candidates were dismissed because of a lack of kinematic consistency. In the 2000s, the search for jet rotation was extended to molecular jets and outflows  with the advance in millimeter and sub-millimeter wavelength interferometry \citep{lee2006, lee2007a}. Using high J transitions of SiO \citep{palau2006} that trace material in the primary jet/wind launched from the disk, \citet{lee2007b, lee2008, lee2009} reported velocity gradients in the HH211 SiO jet and constrained the jet launching radius to within 0.05 AU of the star. However, subsequent observations at a better angular resolution did not confirm these results \citep{lee2010}.

The  challenge of detecting rotation in the optical and molecular jets lies in the fact that the spectral line emission from the jet is broad (20 to 100 \kms-1) due to turbulence and other motions, making it difficult to discern any velocity gradients that are much smaller in magnitude. In addition, asymmetries in shocks can produce velocity gradients that mimic rotation. More recently, velocity gradients were reported in CO outflows \citep[e.g.][]{launhardt2009, zapata2010, zapata2015, chen2016}, demonstrating another promising avenue in the search for jet rotation. However, CO emission may not trace the primary wind due to its low critical density, and is often mixed with the surrounding molecular gas accelerated by the wind. Hence, its interpretation can be complicated by the distribution of the circumstellar material. A more complete review on the subject can be found in \citet{ray2007} and \citet{frank2014}.

The hydrogen recombination masers around the young star MWC349A offer a unique opportunity for studying the structure and dynamics in a disk and wind at an angular resolution unattainable using thermal emission. The target of interest,  
MWC349A, is in a binary system located at a distance from the Sun of 1.2 kpc \citep{cohen1985}. The object is classified as a Herbig B[e] star surrounded by ionized gas seen in hydrogen recombination line emission  \citep{martin1989a} as well as continuum emission at centimeter wavelengths \citep{martin1993, tafoya2004}. Its companion, MWC349B, is offset by $2''.4$. A remarkable and peculiar property of MWC349A is that a subset of the $\alpha$ transition of the hydrogen recombination lines are masing, making it one of a small number of hydrogen recombination maser sources known. The maser emission was first identified in the H29$\alpha$, H30$\alpha$,  and H31$\alpha$ transitions \citep{martin1989a}. Additional maser transitions were discovered subsequently at sub-millimeter wavelengths \citep{thum1994b}, and later at infrared wavelengths \citep{strelnitski1996a,thum1998}. It appears that the $\alpha$ transitions from $7 \le {\rm n} \le 39$ all exhibit maser action \citep{martin1994, thum1998}, while at n $> 39$, the recombination spectra all exhibit a single feature of thermal like Gaussian profiles. High angular resolution observations of the H30$\alpha$  line revealed that the emission occurs at a collection of discrete locations from a disk structure \citep{planesas1992, weintroub2008}. However,  subsequent observations of the H30$\alpha$ transition by \citet{martin2011} found that only those closer to the systemic velocity arise from the disk; masers at more extreme velocities from the systemic velocity are formed in a wind. Using a non-LTE radiative transfer calculation, \citet{rubio2014} modeled the H30$\alpha$  masers and constrained the disk to a nearly edge-on geometry with an inclination angle of $6.5^\circ$ from the line of sight.

Here we present simultaneous observations of the H30$\alpha$ and H26$\alpha$ masers with the Submillimeter Array (SMA) in its Very Extended configuration. The high signal-to-noise data, coupled with an angular resolution as fine as $0''.3$, yield relative astrometry for masers better than 2 milli-arcsecond (mas) precision for masers stronger than 20 Jy. In addition, the first high resolution H26$\alpha$ transition images reveal that the lower quantum number masers are excited in the inner part of the disk as well as in the denser part of the wind than the H30$\alpha$ line masers. We found that masers in the disk are amplified in a thin annulus at a fixed radius for each hydrogen recombination transition. Furthermore, the maser velocity structure in the wind demonstrates Keplerian-like rotation, which implies that a substantial fraction of the angular momentum is extracted by the wind. The paper is organized as follows: In Section 2, we present the observational setup and the details of data processing. In Section 3, we present the observational results. In Section 4, we discuss the implication of the results. A brief conclusion is provided in Section 5.

\section{Observations and Data Calibration}

Observations of MWC349A were carried out on 2012 October 12 using the SMA\footnote{The SMA is a joint project between the Smithsonian
Astrophysical Observatory and the Academia Sinica Institute of Astronomy and
Astrophysics, and is funded by the Smithsonian Institution and the Academia
Sinica.} \citep{ho2004} in the Very Extended configuration. Six antennas were operational during the observations, providing projected baselines from 84 to 509 m. Observations made use of both 230 GHz and 400 GHz receivers simultaneously, targeting the H${30 \alpha}$ and H${26 \alpha}$ lines at  rest frequencies of 231.9009 GHz and 353.6228 GHz, respectively. The digital correlator was configured to a resolution of 256 channels per 104-MHz window for three contiguous windows close to the recombination lines. For this configuration, the channel spacing is 0.41 MHz, corresponding to 0.50 and 0.35 \kms-1\ at 232 and 354 GHz, respectively. The remaining windows in the 2-GHz band were set to 32 channels per 104-MHz band. The zenith opacity reported from the CSO $\tau$-meter was  0.1 at 225 GHz. The double side band system temperatures varied from 80 to 150 K for the 230 GHz receivers, and 230 to 600 K for the 400 GHz receivers. The atmospheric phases were stable during the start of the track, but deteriorated after 6 hours into the observations.


{The phase center for MWC349A during the observations was $\alpha_{\mathrm{J}2000}$=$20^{\mathrm{h}}32^{\mathrm{m}}45.53^{\mathrm{s}}$,
$\delta_{\mathrm{J}2000}$=$40^{\circ}39^{\prime}36.61^{\prime\prime}$. QSO J2015+371, at an angular distance of $4.7^\circ$ from MWC349A, was observed periodically to monitor gain variations during the course of the observations.} In addition, Callisto was observed for flux calibration and 3C84 for passband calibration. Since accurate passband calibration is essential for determining the astrometry of maser spots across the spectral line, we first observed 3C84 for 2 hours in the dual receiver mode with the lower-frequency IF tuned to the H30$\alpha$ line and the high-frequency IF tuned to the H26$\alpha$ line. Additional passband data were taken for the high-frequency IF by rerouting the lower-frequency signal through the high-frequency IF path to characterize the passband response. This approach enabled us to obtain higher signal-to-noise passband data for the H26$\alpha$ line since 3C84 is brighter and system temperatures are lower in the 230 GHz band.

We performed data calibration using the IDL subset MIR software package. The absolute flux was bootstrapped to Callisto, which yields a flux density of 2.3 Jy at 230 GHz and 1.6 Jy at 345 GHz for QSO J2015+371. The continuum flux densities derived for MWC349A are 1.7 Jy at 230 GHz and 2.4 Jy at 345 GHz, in agreement with the model of an ionized wind source \citep[e.g.,][]{tafoya2004}.
The time-dependent gain calibration is achieved through  the nearby QSO J2015+371.  
For passband calibration of the 3 windows with 256 channels each, we averaged every 16 channels (equivalent to 6.5 MHz) of the 3C84 data before solving for antenna based passband solutions. The smoothing reduces the noise contribution from the passband solution to the spectral line data of MWC349A (see Section 3). In this approach,  passband variations on scales less than  6.5 MHz are smoothed out.
We believe that there is no systematic phase structure on frequency scales smaller than 6.5 MHz. Figure 1 presents visibility phase of the baseline involving antennas 5 and 6 for the H30$\alpha$ and H26$\alpha$ lines. As shown in the figure, visibility phases in the velocity range from 25 to 42 \kms-1\ have no apparent phase structure within 16 channels.
After applying passband solutions,  the phase residual in the higher resolution spectral channels of 3C84 is 2$^\circ$. For the lower spectral resolution windows, no average was performed when deriving the passband solution. 

The calibrated visibility data of MWC349A were exported to the MIRIAD format. Since the continuum emission of MWC349A is unresolved \citep[expected to be $0''.044$ at 232 GHz and $0''.032$ at 354 GHz according to the model of][]{tafoya2004} in the Very Extended array configuration, we performed self calibration on the continuum data  using a time interval of 30 sec to further remove phase fluctuations at shorter time scales. The self calibrated spectral line visibility data were then Fourier transformed and the images were cleaned. For the H26$\alpha$ line, the continuum data did not have sufficient signal-to-noise ratios for self calibration at time intervals of several minutes. Hence, we resorted to using the strongest maser feature at $30$ \kms-1\ for self calibration. Since the self calibration of the H30$\alpha$ and H26$\alpha$ lines adopted different references, images of the H26$\alpha$ and H30$\alpha$ emission are not registered in position. The synthesized beam is $0.36'' \times 0.31''$ for the H26$\alpha$ line and $0''.54\times 0.45''$ for the H30$\alpha$ line. The $1 \sigma$ rms noise in the self calibrated images is 80 mJy and 110 mJy per 0.5 \kms-1\ channel in the H30$\alpha$ and H26$\alpha$ lines, respectively.

\section{Results}

Figure 2 presents the spectra of the H30$\alpha$ and H26$\alpha$ lines. The spectra are color coded by the \vlsr\ velocity of the channel. Both recombination line spectra show two emission peaks near $-18$ and 31 \kms-1, respectively. The redshifted peak is a factor of 2 stronger than the blueshifted peak. In between the two peaks lies a flat portion of the spectrum detected at a level of more than 5 Jy. 

Maser spots arise from  compact regions that are not spatially resolved by the SMA observations at resolutions of $0.3'' - 0.5''$. If visibility phases are well determined across the spectral line, centroid fitting to the maser emission yields astrometry with precisions better than the spatial resolution. As shown in Figure 1, over the spectral window of 100 \kms-1, the visibility phase varies from $-10^\circ$ at $-20$ \kms-1\ to $5^\circ$ at 35 \kms-1 for the H30$\alpha$ line, indicating a systematic spatial shift in the maser position. The ability to measure the position is limited by the phase noise in the spectrum, which in turn is related to the signal-to-noise ratio (SNR) in the data.
The position error, $\Delta \theta_{\mathrm{fit}}$ is related to the size of the synthesized beam and  the SNR through the approximate relation
$$ \Delta \theta_{\mathrm{fit}} =  {\theta_\mathrm{{beam}} \over {2~{\rm SNR}}} $$ \citep{condon1997}. 
The synthesized beam size at 354 GHz for the H26$\alpha$ line observations was $0.36'' \times 0.31''$. We calculated a noise value of 110 mJy by measuring the standard deviation in a 0.5 \kms-1\ wide channel that does not contain significant maser emission. For H26$\alpha$ masers with flux intensities of 40 Jy, the corresponding position error was 0.8 mas. The synthesized beam at 232 GHz for  the H30$\alpha$ observations was $0.''54 \times 0''.45$. The $1 \sigma$ rms in the channel of width 0.5 \kms-1\ was 80 mJy, which gave a position error of 0.6 mas for maser features of 40 Jy.

In addition to statistical errors from centroid fitting, the phase noise in the passband data introduces additional error to the maser position through passband calibrations. The passband calibrator 3C84 had a flux density of 9.5 Jy at the 230 GHz band during the time of the observations. If one ignores the slight difference in the on-source time between the 3C84 and MWC349A observations, the noise contribution from the passband is equivalent to that of a flux density of 9.5 Jy. We improve the SNR in the passband solution by a factor of 4 by applying a smoothing of 16 channels when deriving antenna based passband solutions.
After the channel averaging, the  phase noise in 3C84 is reduced to $\Delta \phi$ = 2$^\circ$. This phase value is equivalent to a position error, $ \Delta \theta_{\mathrm{BP}} =  \theta_{\mathrm{beam}}  {\Delta \phi \over { 360^\circ}}, $
of 2.5 mas and 1.8 mas for  H30$\alpha$ and H26$\alpha$, respectively.
The total error $\Delta \theta$ in each maser position is given by $\sqrt{\Delta \theta_{\mathrm{fit}}^2 + \Delta \theta_{\mathrm{BP}}^2}$.

Figure 3 shows the spatial distribution of the H26$\alpha$ and H30$\alpha$ spots determined from the centroiding method. Here we only plot the data in the velocity range of $-29$ to 47 \kms-1\ because of signal-to-noise considerations. The error bar corresponds to the $\pm 1 \sigma$ error given by $\Delta \theta$.  Masers from $-12$ to 25 \kms-1\ for both transitions are mostly distributed in a linear structure in the southeast-northwest direction. The least-squares fit to the maser distribution within this velocity range yields a position angle of $101.9^\circ \pm 0.8^\circ$ for the H30$\alpha$ emission and $98^\circ \pm 1^\circ$ for the H26$\alpha$ emission. The position angles of the two maser distributions are consistent within the statistical error in the fit. The position angle of the radio continuum emission is $8^\circ$ \citep{martin1993, tafoya2004}. The position angle of the maser structure is consistent with the orientation of the disk inferred  from the centimeter-wavelength  continuum and the previous high resolution data of H30$\alpha$ obtained from the SMA and PdBI \citep{weintroub2008, martin2011}.

As shown in Figure 3, the maser emission exhibits a velocity gradient from southeast to northwest. Masers are red shifted toward the southeast and are blue shifted toward the northwest. The sense of velocity gradients is the same for the two lines. 
The spatial extent of the H26$\alpha$ maser spots in the disk is slightly smaller than that of the H30$\alpha$ maser spots in the disk.  This implies that H26$\alpha$ masers arise from zones closer to the center of the disk than the corresponding H30$\alpha$ maser spots. 

In addition to masers distributed along the linear structure, those with velocities greater than about  20 \kms-1\ from the systemic velocity are distributed off the disk plane. This is most apparent in the  H26$\alpha$  masers at \vlsr\ velocities from  30 to 60 \kms-1\ and $-20$ to $-40$ \kms-1. It has  been suggested that these masers arise from the  ionized wind of gas lifted off the disk plane \citep{martin2011}. 

To compare the spatial distribution and kinematics between the two maser transitions, their relative positions must be aligned to an accuracy of 2 mas, the uncertainty limited by the passband calibration.
As seen in Figure 3, there is a relative spatial shift between the H30$\alpha$ and H26$\alpha$ masers due to the self calibration that referenced different sources (see Section 2). 
We explored several avenues to align the two maser distributions. The first approach is to align the 230 GHz and 345 GHz continuum emission. Since the 230GHz and 345 GHz continuum emission arise from the free-free emission of the ionized gas and are spatially unresolved in the SMA observations, one expects that they share the same position. We fit the two continuum images. The fit to the 230 GHz continuum data yields an offset  ($\Delta {\alpha}, \Delta {\delta}$) = ($0''.000 \pm 0.001, 0''.000  \pm 0.001$), as expected from the data after self calibration using a point source model at the phase center. For the 345 GHz continuum, reliable fits were not attainable at all due to residual phase variations despite the self calibration. Therefore, we resorted to the second approach by using the strongest blueshifted and redshifted features in the H30$\alpha$ and H26$\alpha$ masers in the disk. This approach assumes that two maser transitions in the disk share a common geometric center (see discussions in Section 4.1).  We use strongest masers in the disk for alignment since they have the most  reliable positions due to their high SNR. In addition, they arise from the tangential points in the disk (see discussions in Section 4.1). The centers derived this way are ($\Delta \alpha, \Delta \delta$) = ($-0''.001 \pm 0''.002, +0''.001 \pm 0''.002$) for the H30$\alpha$ masers and ($-0''.019 \pm 0''.002, +0''.002 \pm 0''.002$)  for the H26$\alpha$ masers, respectively. These centers are marked by the large triangles in Figure 3.

To further analyze the kinematics, we  present rotation curves for both maser transitions in Figure 4.  A rotation curve is the line-of-sight velocity against projected distance $l$, defined as the separation between masers and the center of the disk projected on the plane of the sky.
The rotation curves of  the H30$\alpha$ and H26$\alpha$ masers exhibit a linear relation in the position-velocity diagram between $-12$ to 25 \kms-1. Maser velocities outside of this range decrease at increasing $l$. More significantly, the linear portion of the rotation curve shows different slopes between the maser transitions, with  the H26$\alpha$ masers exhibiting a steeper gradient than those of the H30$\alpha$.

\section{Discussion}

\subsection{Maser Geometry}

Based on the spatial distribution and the kinematics shown in Figures 3 and 4, we divide the masers in two groups: Group {\it I} are those close to the cloud systemic velocity from $-12$ to 25 \kms-1; Group {\it II} are those at the blueshifted velocities of \vlsr $<$ $-15$ \kms-1\ and at the redshifted velocities of \vlsr $>$ 30 \kms-1.

The Group {\it I} maser features between $-12$ to 25 \kms-1\ are distributed in a linear strip in the southeast-northwest direction. They consist of emission at the flat part of the spectra as well as those with increasing intensities toward the blueshifted and redshifted velocities (see Figure 2). In addition, these masers lie in the linear portion of the rotation curve. This group of masers  likely arises from the disk at a fixed radius of $R$ as proposed by  \citet{weintroub2008} based on the study of the H30$\alpha$ masers. 
The position angle of the maser disk is perpendicular to the position angle of the wind  in \citet{martin1993} and \citet{tafoya2004}, respectively.

The H26$\alpha$ Group {\it I} masers are distributed in a linear extent smaller than that of the H30$\alpha$ masers. In addition, the former maser transition exhibits a larger velocity gradient. Both findings indicate that H26$\alpha$ masers are formed inside the radius of the H30$\alpha$ masers.  This is consistent with the fact that the formation of the H26$\alpha$ masers requires higher densities \citep{strelnitski1996b}, a condition that is met in a centrally condensed disk. 

These linear relations suggest a simple model in Figure 5: an edge-on disk with masers in each transition distributed in a thin annulus at a fixed radius from  $\phi = 0^\circ$ to $\pm 90^\circ$. Here $\phi$ is the angle between the line of sight and the vector connecting the star and the maser. It is conceivable that masers populate the entire annulus. However, the continuum emission from the disk within the annulus can be optically thick \citep{ponomarev1994}. Thus, it attenuates the maser emission in the section of the annulus from $\phi = \pm 90^\circ$ to $180^\circ$ in an edge-on disk.

With this geometry,  the line-of-sight velocity  with respect to the systemic velocity, $V_{\mathrm{los}}$,  is given by
\begin{equation}
  V_{\mathrm{los}}=V_{\mathrm{orbit}} ~\mathrm{\sin}\phi,
\end{equation}
where $V_\mathrm{orbit}$ is the circular velocity of the masers. Such a geometry leads to 
\begin{equation}
 V_{\mathrm{los}} = \sqrt{GM \over R^3} l
\end{equation}
for Keplerian motion. Here $M$ is the enclosed mass, and the projected distance $l = R~\mathrm{\sin}\phi$.
This relation explains the constant velocity gradient in Figure 4 for masers between $-12$ to 25 \kms-1. Masers with \vlsr\ close to the systemic velocity lie in the section closer to the line of sight between the observer and the star (with small angles of $\phi$). Masers at the more blueshifted and redshifted ends in this group are close to the tangent point in the annulus (with $\phi \sim \pm 90^\circ$).

The velocity gradient of the Group {\it I} masers in the disk can constrain the enclosed mass within the orbit. A  least-squares fit to the rotation curves yields a velocity gradient $V_\mathrm{los}/l$ of $690 \pm 10 $ \kms-1\ per arcsecond, and a systemic velocity $V_0$ of $7.11 \pm 0.01$ \kms-1\ for the H30$\alpha$ masers, and $890 \pm 20$ \kms-1\ per arcsecond and a systemic velocity of $7.33 \pm 0.01$ \kms-1\ for the H26$\alpha$ masers, respectively. The systemic velocities of the two maser transitions are consistent within the error bar, and are close to the mean velocities between the blueshifted and redshifted maser peaks \citep{thum1994b}. We adopt the average velocity of 7.2 \kms-1\ as the systemic velocity of the system.

Eq. [2] shows that the ratio of annular radii of the maser disks at 232 and 354 GHz is
$R_{232}/R_{354} \sim 1.18 \pm 0.04$. This is one of the more robust quantitative results of this study. For discussion purposes we can convert this ratio into a power law form, $R \sim \nu^{-0.42 \pm 0.10}$.
This dependence follows the prediction of non-LTE maser emission in an environment where the density decreases with radius.

\citet{strelnitski1996b}  made detailed maser models for MWC349 based on the population departure coefficient calculations of \citet{storey1995}. They found that the maser emission from transitions with upper quantum number $n_{\rm max}$ is fairly sensitive to density $n_{\rm e}$  and follows the relation $n_{\rm max} \sim n_{\rm e}^{-0.18}$. \citet{thum1994b} followed a somewhat different methodology and determined that $n_{\rm max} \sim n_{\rm e}^{-0.22}$. If we adopt the scaling law $n_{\rm max} \sim n_{\rm e}^{-0.20 \pm 0.02}$, we can use our measurement to derive a power law dependence for $n_{\rm e}$ as follows. Since the recombination line frequency scales as $\nu^{-3}$, our results indicate that the radius-frequency relation $R\sim  \nu^{-0.42 \pm 0.10}$ becomes $R \sim n_{\rm e}^{-0.25 \pm 0.06}$ or
$n_{\rm e} \sim R^{-4 \pm 1}$. Hence our result is consistent with non-LTE maser theory and a radially
declining density profile which is often found in thin accretion disks \citep[e.g.,][]{pringle1981}. 

We now turn our attention to the estimation of the central stellar mass of MWC349A  from 
the aligned rotation curves shown in Figure 6. In this figure the systemic velocity of 7.2 \kms-1\ has been removed and the fits to the linear parts of the rotation curve shown. We believe that these rotation curves trace the emission to the limiting azimuth angles from $0^\circ$ to $\pm 90^\circ$ for the reasons described in Section 4.2. However the exact separation between the wind and disk masers is not sharply defined. We judge that a mass of 10 \msun, which is appropriate for a maximum projected diameter of 51 and 42 mas (see Figure 6) for the H30$\alpha$ and H26$\alpha$ emission is an appropriate determination. It is possible that the masers with velocities up to 19 and 21 \kms-1\ should be included in the disk, in which case the central mass would be 15 \msun. This mass range of 10 to 15 \msun\ is significantly smaller than previous estimates of 25 to 35 \msun  \citep[e.g.,][]{thum1994b,ponomarev1994}
which do not make a distinction between the disk masers that are in true Keplerian motion, and wind masers that trace gas not gravitationally bound to the star.

Similar to the Group {\it I} masers, the Group {\it II} masers present distinct spatial and kinematic characteristics as well. Maser emission at \vlsr $>$ 30 \kms-1\ is found in a linear structure off the plane of the disk with a noticeable spatial offset between the two transitions. Both H30$\alpha$ and H26$\alpha$ maser features are distributed in an approximately linear structure extending 0.02$''$ to the northwest at a position angle of  $-20^\circ$ and $-9^\circ$, respectively (see Figure 3).  The maser features at blueshifted \vlsr $<$ $-17$ \kms-1\  are also offset from the disk plane for the H26$\alpha$ line, although the extent is relatively smaller than the redshifted masers. Furthermore, these masers are located outside of the linear part of the rotation curves (see Figure 6). These properties suggest that Group {\it II} masers  arise from the wind as proposed by \citet{martin2011}. The strongest masers in this group are likely excited along the tangent point in the wind bubble where a longer path of amplification is achieved (see discussion in Section 4.2). The fact that these masers do not follow a Keplerian rotation is consistent with the notion that they are not dynamically bound to the star.

\subsection{Maser Amplification}

Hydrogen recombination maser emission in MWC349A exhibits a double-peaked profile. Assuming that masers are excited in a disk with an inner hole, \citet{ponomarev1994} reproduced the double-peaked profile when the inner radius is greater than 70\% of the outer radius. Our observations confirm that masers from the same transition are indeed excited in a thin annulus of a fixed radius. This geometry shown in Figure 5 can explain why the Group {\it I} maser emission from the $\phi = \pm 90^\circ$ points in the disk is much stronger than that from near $\phi = 0^\circ$. Our analysis is based on a  simplistic model rather than on  a detailed radiative transfer calculation in \citet{ponomarev1994}. The velocity along the line of sight near $\phi = \pm 90^\circ$ is given by
\begin{equation}
 V_{\mathrm{los}}=V_{\mathrm{orbit}} ~\mathrm{\cos}\phi^\prime = \sqrt{GM \over R^3} ( 1 -  {\phi^\prime}^2),
\end{equation}
where $\phi^\prime$ is $\phi - 90^\circ$. Note that the velocity gradient is only zero at $\phi^\prime = 0^\circ$. We define a coherent amplification bandwidth $V_{\mathrm{c}}$, which gives the azimuthal
angle range contributing to the maser line formation of 
\begin{equation}
  \Delta \phi = 2\sqrt{2 V_ \mathrm{c} \over V_{\mathrm{orbit}}}.
\end{equation}
The coherent amplification length is then $R \Delta \phi$, or
\begin{equation}
  L( \phi = 90^\circ) = 2R \sqrt{2 V_ \mathrm{c} \over V_{\mathrm{orbit}}}.
\end{equation}

For the H30$\alpha$ transition we have $V_\mathrm{orbit}$ = 17.2 \kms-1\ at  $R$ = 25 mas (30 AU).   If we adopt $V_{\rm c}$ = 0.5 \kms-1\ then $\Delta \phi$ = 0.48 radians or 27$^\circ$. The coherent amplification length is therefore 12 mas (14 AU). At $\phi = 0^\circ$ the velocity gradient is zero for all radii, but the population inversion is highly sensitive to the density which decreases monotonically with $R$. Since our model for the H26$\alpha$ emission has $R$ = 21 mas (25 AU), we assume that the peak shifts by about 1.0 mas per quantum number, $i.e.$ (25-21)/4 = 1.0 mas (1.2 AU). We further assume that the coherent amplification length is approximately equal to twice this value or 2 mas (2.4 AU) The amplification length ratio, $r = L(\phi = 90^\circ)/L(\phi = 0^\circ)$, is 12/2 or about 6.  Depending on details of the maser model the maser intensity scales exponentially with $L$ for unsaturated amplification, $L^3$ for saturated amplification in a filamentary structure, and $L$ for saturated amplification in a one dimensional  geometry \citep{goldreich1972}. For these three cases the intensity ratio are expected to be 403 ($e^6$), 216 ($6^3$), and 6, respectively. The observed intensity ratio for our experiment is about 16 and 8 for the redshifted and blueshifted lines, respectively. Historically, these ratios vary over the range 5-30 \citep{gordon2001}. This range fits within the range of our simple model, perhaps favoring the one-dimensional saturated case. This situation differs from the case of water megamasers where the annular emission region is thick and masers near the systemic velocity  dominate because of a larger coherent length \citep[e.g.][]{braatz1996}.

While the spatial characteristics of the masers in the disk seem to be fairly stable over the time scale of several years based on the disk maser distributions observed in 2004, 2005 and 2012 \citep[][and this paper]{weintroub2008, martin2011}, the masers in the wind seem to be more variable and sensitive to excitation conditions. We note that the redshifted wind features near 30 \kms-1\ extend below the disk in the H30$\alpha$ image (see Figure 3), while they extend above the disk in the H26$\alpha$ image. This suggests that maser amplification may be sensitive to density and other excitation conditions. In addition, in the 2005 data by \citet{martin2011} the H30$\alpha$ feature extends above the disk, suggesting temporal variability. We note that the intensity profile monitored over many years by \citet{gordon2001} shows considerable variation, which indicates changing excitation conditions.

\subsection{Angular Momentum in the Wind}

The angular momentum in stellar systems like the Sun is four orders of magnitude lower than that in dense cores. Therefore, angular momentum has to be redistributed during the collapse of dense cores and formation of protostars. While the formation of binaries and multiples can be effective in redistributing angular momentum \citep[e.g.][]{chen2013}, disks must shed sufficient angular momentum through the wind and the jet in order for the gas to be accreted onto the protostar. The wind is launched via a magneto hydrodynamical process from the disk and then interacts with the ambient material to give rise to outflows seen in spectral lines and continuum in the optical, infrared and radio wavelengths. The two models on protostellar outflows, i.e. the X-wind \citep{shu2000, shang2007} and the disk wind \citep{konigl2000}, differ significantly in zones where the wind is launched. The X-wind model has the wind launched in the co-rotation radius between the protostar and the disk at a radius of $< 0.1$ AU, while the disk wind model features the launch radius of several AU. Since the corresponding angular scales of the jet launching radii are $0.3 - 30$ mas even for the nearby star formation regions such as the Taurus molecular cloud, they are beyond the sensitivity/resolution of many telescope facilities. Observations of jet rotation that occurs at much larger spatial scales offer the added benefit of constraining the jet launching zones and distinguishing the two outflow models.

In an attempt to confirm that protostellar jets carry out angular momentum, considerable observational effort has been devoted to searching for rotation in protostellar jets and outflows using thermal emission \citep{chen2016}. Masers seen in jets and outflows complement the thermal probes thanks to their brightness that enables much higher angular resolution observations. \citet{burns2015} presented multi-epoch observations of an H$_2$O maser jet in the massive protostar S235AB-MIR using the VLBI network VERA. They found a velocity gradient and proper motion in masers that are consistent with jet rotation. The hydrogen recombination masers in MWC349A offer the direct probe of the ionized wind and its kinematics. It appears that the Group {\it II} masers with \vlsr\ close to $-15$ and 30 \kms-1\ have projected line-of-sight velocities similar to a Keplerian relation. Masers with \vlsr\ much less than $-15$ \kms-1\ or much greater than 30 \kms-1\ are fainter, thus do not have sufficient SNR to determine their positions accurately. 

The radio continuum observations of MWC349A yield electron densities in the wind, thus providing a direct measure of the mass loss in the ionized gas. \citet{tafoya2004} carried out centimeter-wavelength continuum observations and estimated a mass loss of $5 \times 10^{-6}$ \msun~yr$^{-1}$. Stars in the protostellar phase of their evolution are expected to experience much higher rates of accretion and mass loss than stars in the more evolved phase of their evolution. For an order of magnitude estimate, we assume a constant mass loss rate over $5 \times 10^5$ yrs, the accretion time scale of a massive protostar \citep{zhang2015}. Then, the lower limit of the total mass loss over the lifetime of its accretion  amounts to 2.5 \msun. The present disk mass of MWC349A is constrained to $< 5.7$ \msun\ based on the near IR observations \citep{danchi2001}. It is reasonable to expect that the star is surrounded by a more massive disk during the protostellar stage \citep{cesaroni2007}. Nevertheless, the lower limit of the total mass loss during the entire accretion history is a significant fraction of the disk mass. Therefore, the wind removes significant angular momentum during the accretion phase of the star formation. 

The spatial distribution of the maser spots also indicates that the wind is launched at a radius as far out as 25 AU, much larger than models of  the X-wind \citep{shu2000} or the disk wind \citep{konigl2000}. Since the maser emission may not fully trace the distribution of the mass, it is not certain how the material is distributed in the wind. While the masers are distributed over a radius of 25 AU, it remains possible that the majority of the mass ejection occurs at much smaller radii close to the (proto)star. Observations of recombination masers of lower quantum numbers will help define the wind structure in the higher density regime.

\section{Conclusion}

The high signal-to-noise ratio data obtained with the SMA at sub-arcsecond resolutions provide precise relative astrometry of H30$\alpha$ and H26$\alpha$ masers in MWC349A. We found

(1) two groups of masers: Group {\it I} masers  with \vlsr\ close to the systemic velocity from $-12$ to 25 \kms-1\ lie in a disk; and Group {\it II} masers outside of the velocity range are distributed off the disk in a wind;

(2) The H26$\alpha$ masers are distributed in a region inside that of the H30$\alpha$ masers, indicating that the former are excited at higher density regimes in the disk and the wind. The kinematics of masers in the disk indicate that they form in a thin annulus of a fixed radius for each maser transition;

(3) The kinematics in the masers reveal a  wind in the young stellar system that rotates in the same sense as the disk rotation. MWC349A offers strong observational support  that the wind extracts angular momentum from the disk. In doing so, it facilitates gas accretion toward the star through the disk. 

\acknowledgements
We thank K. Young for his help in setting up SMA observations and J. Weintroub for discussions during the data analysis. Data presented in this paper were taken during the Astronomy 191 course in experimental astrophysics at Harvard University through the Director Discretionary Time. We thank R. Blundell for his support of the project.  We appreciate the comments of the anonymous referee, which helped improve the clarity of the paper.

\clearpage

\begin{figure}[ht!]
\includegraphics[width=15cm]{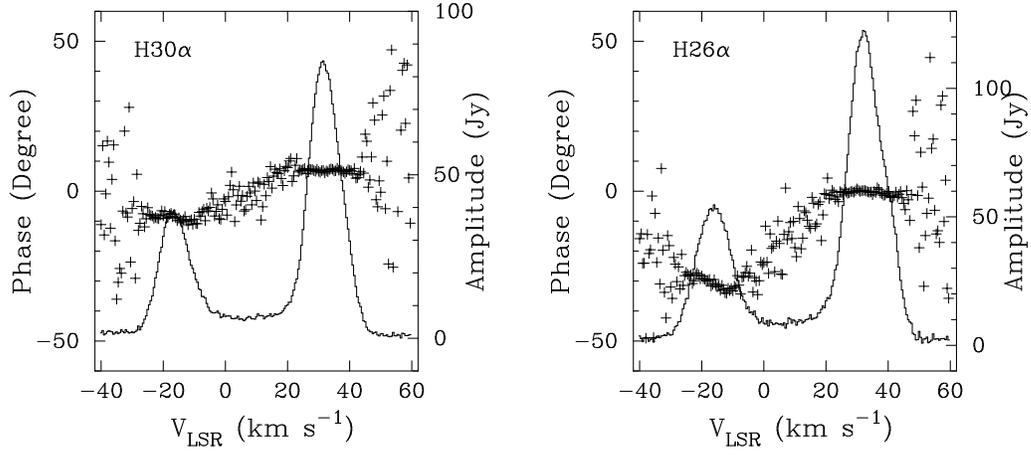} 
\caption{
Visibility phase and amplitude of the  H30$\alpha$  and H26$\alpha$ lines for a baseline of 420 m.
The phase variations across the spectral window are due to the differences in position of the maser spots. The nominal phase sensitivity is 1.2$^\circ$ mas$^{-1}$ and 1.7$^\circ$ mas$^{-1}$ at the frequencies of the H30$\alpha$  and H26$\alpha$ transitions, respectively.}
\end{figure}

\begin{figure}[h]
\begin{tabular}{ p{8.5cm} p{8.5cm} }
\includegraphics[width=8.5cm]{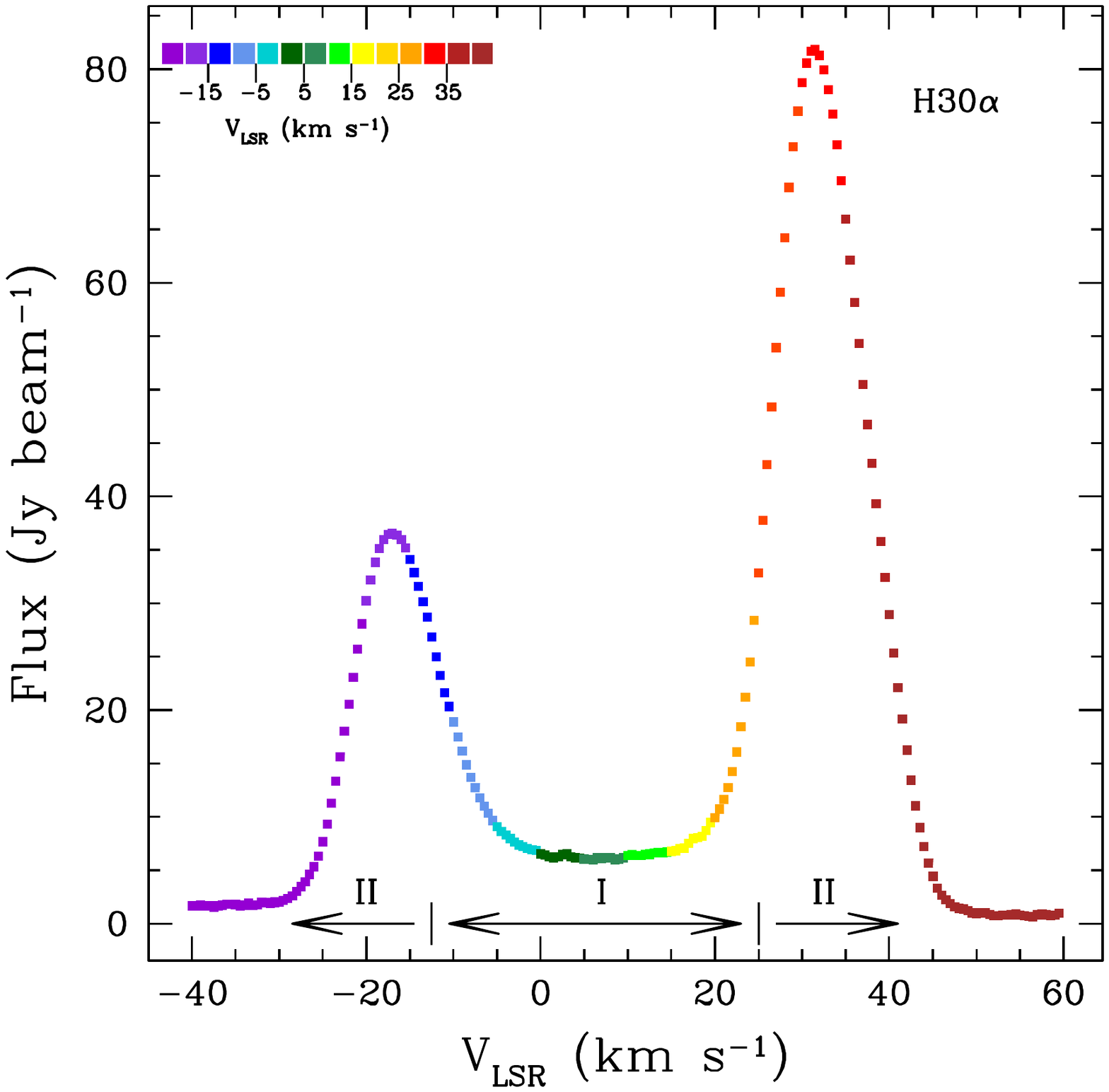} & \includegraphics[width=8.5cm]{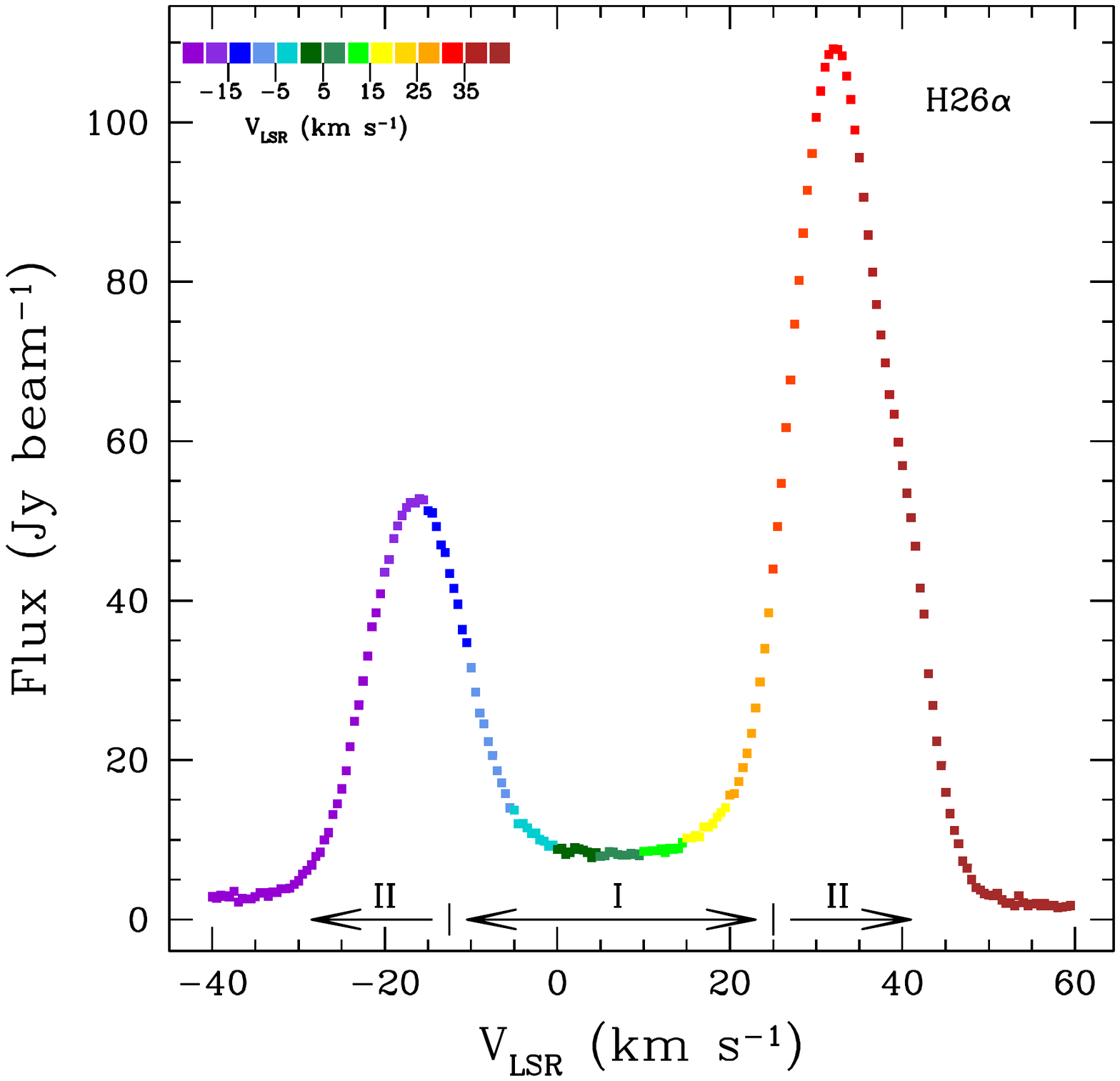} \\
\end{tabular}
\caption{
Line profiles of H30$\alpha$ (231.9 GHz) and H26$\alpha$ (353.6 GHz) masers of MWC349A. The spectra are color coded by the velocity of the channel. The two groups of masers (see discussion in Section 4.1) are marked in the figure by {\it I} for Keplerian disk masers and {\it II} for wind phase masers.
}
\end{figure}

\begin{figure}[h]
\begin{tabular}{ p{8.5cm} p{8.5cm} }
\includegraphics[width=8.5cm]{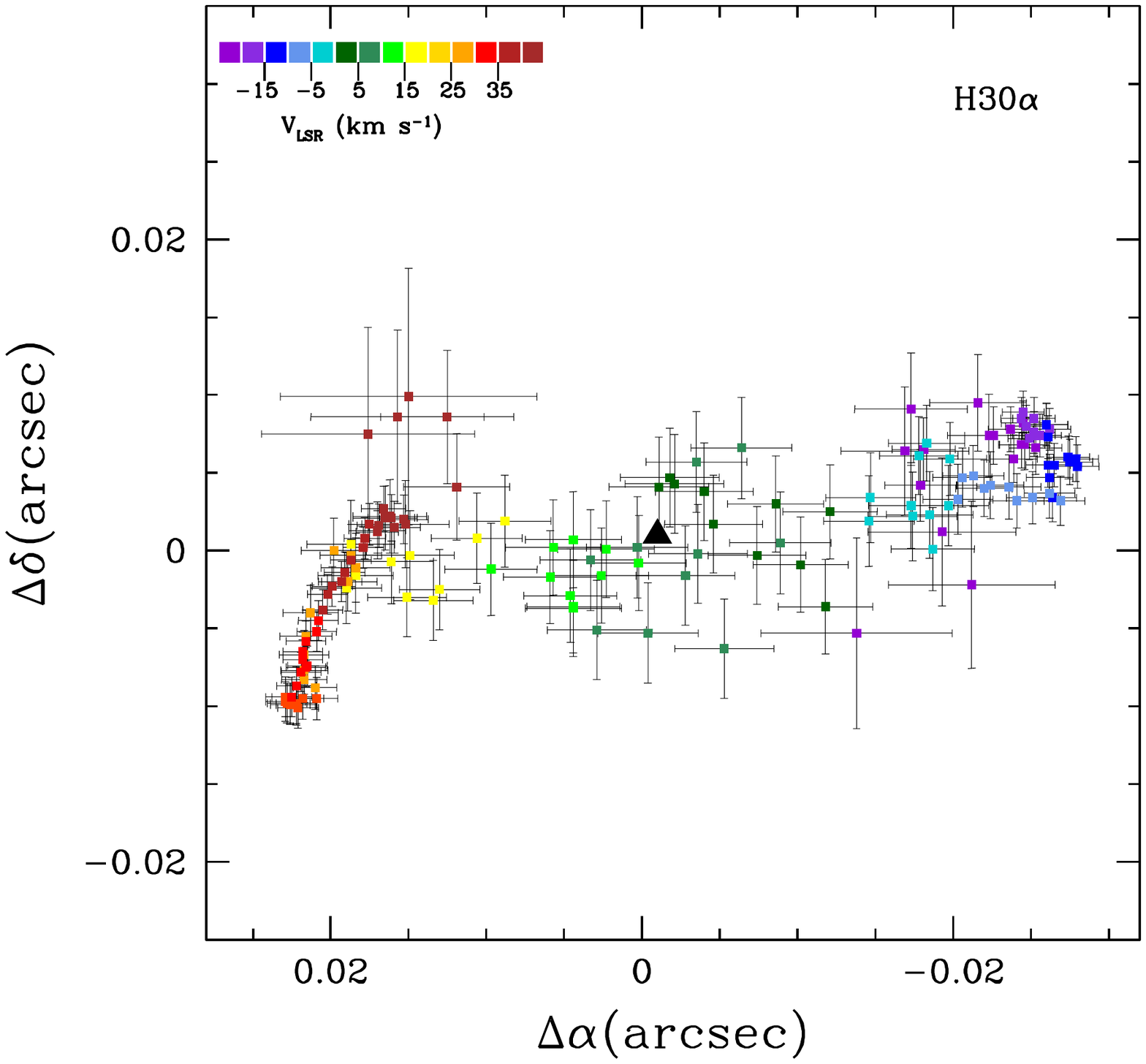} & \includegraphics[width=8.5cm]{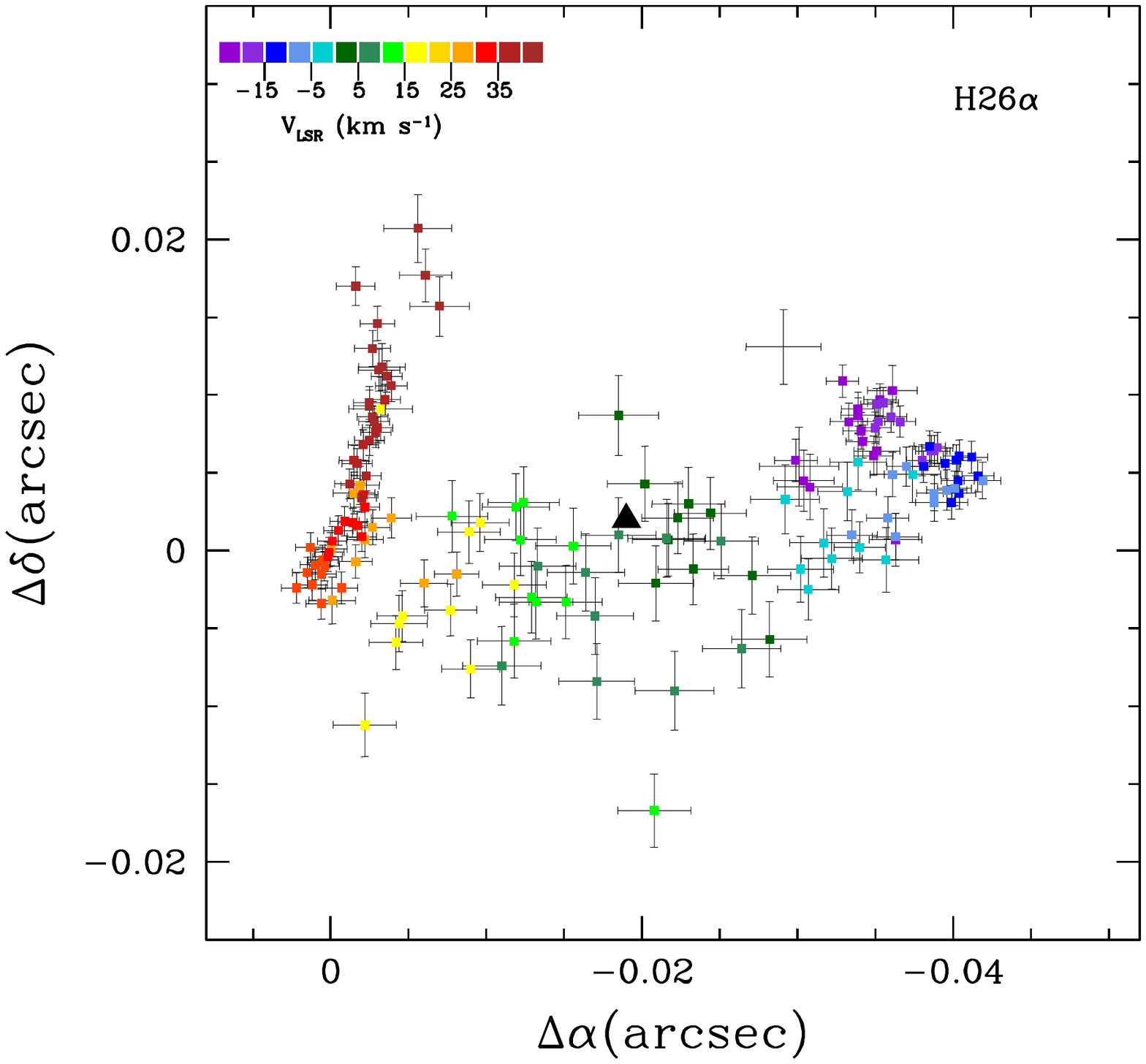} \\
\end{tabular}
\caption{
Spatial distributions of H30$\alpha$  and H26$\alpha$ maser spots represented by square symbols. The color coding marks the \vlsr\ of the maser feature. The cross bars denote the $\pm 1\sigma$ position error from centroid fitting and passband calibration. The H30$\alpha$ image is referenced to the continuum emission at 230 GHz, and the H26$\alpha$ is referenced to the position of its 30 \kms-1 maser feature. The large triangles mark the position of the geometric center of the H30$\alpha$ and H26$\alpha$ masers in the disk, respectively. An angular size of $0''.01$ corresponds to a projected scale of 12 AU.} 
\end{figure}

\begin{figure}[h]
\begin{tabular}{ p{8.5cm} p{8.5cm} }
\includegraphics[width=8.5cm]{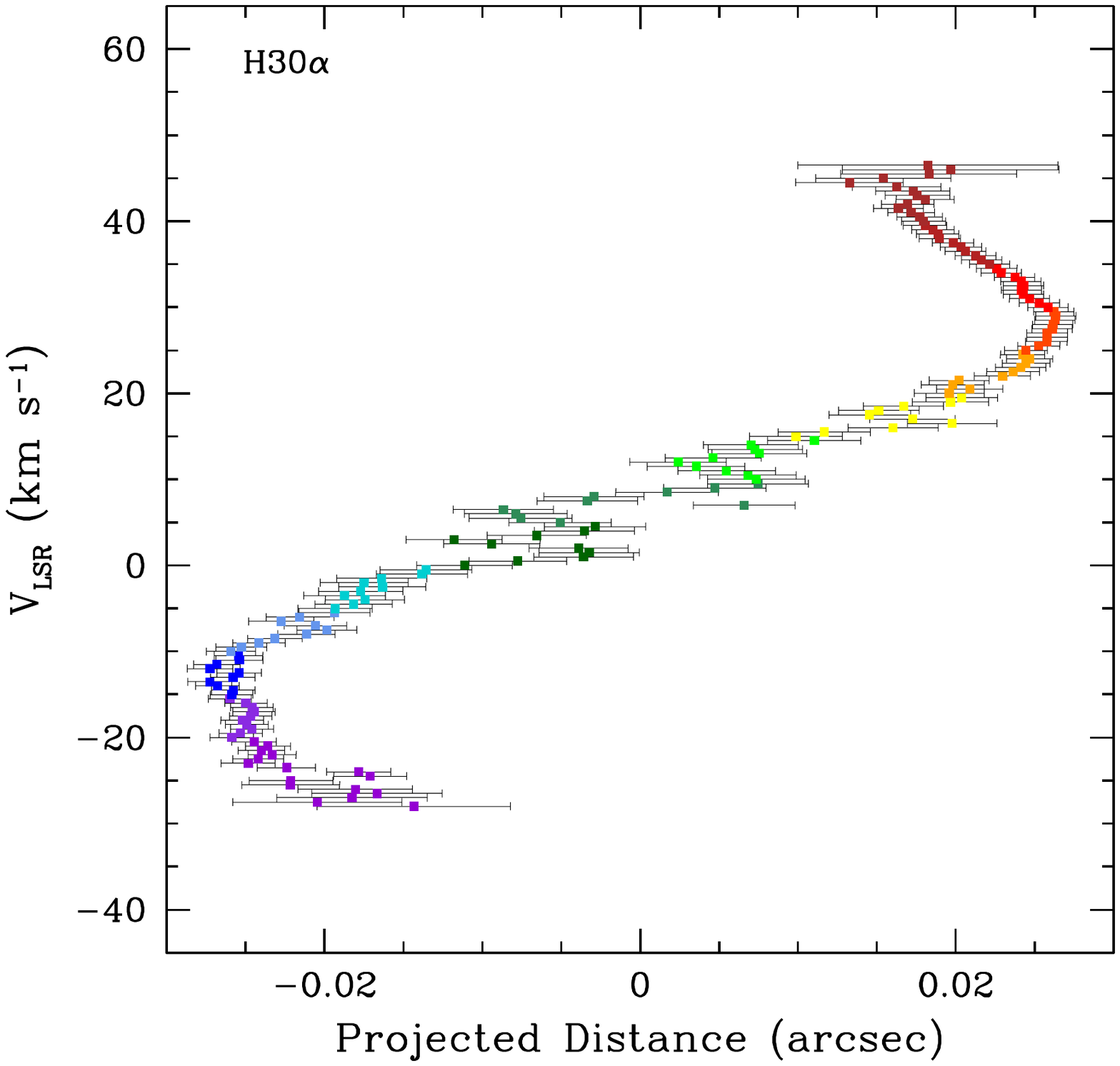} & \includegraphics[width=8.5cm]{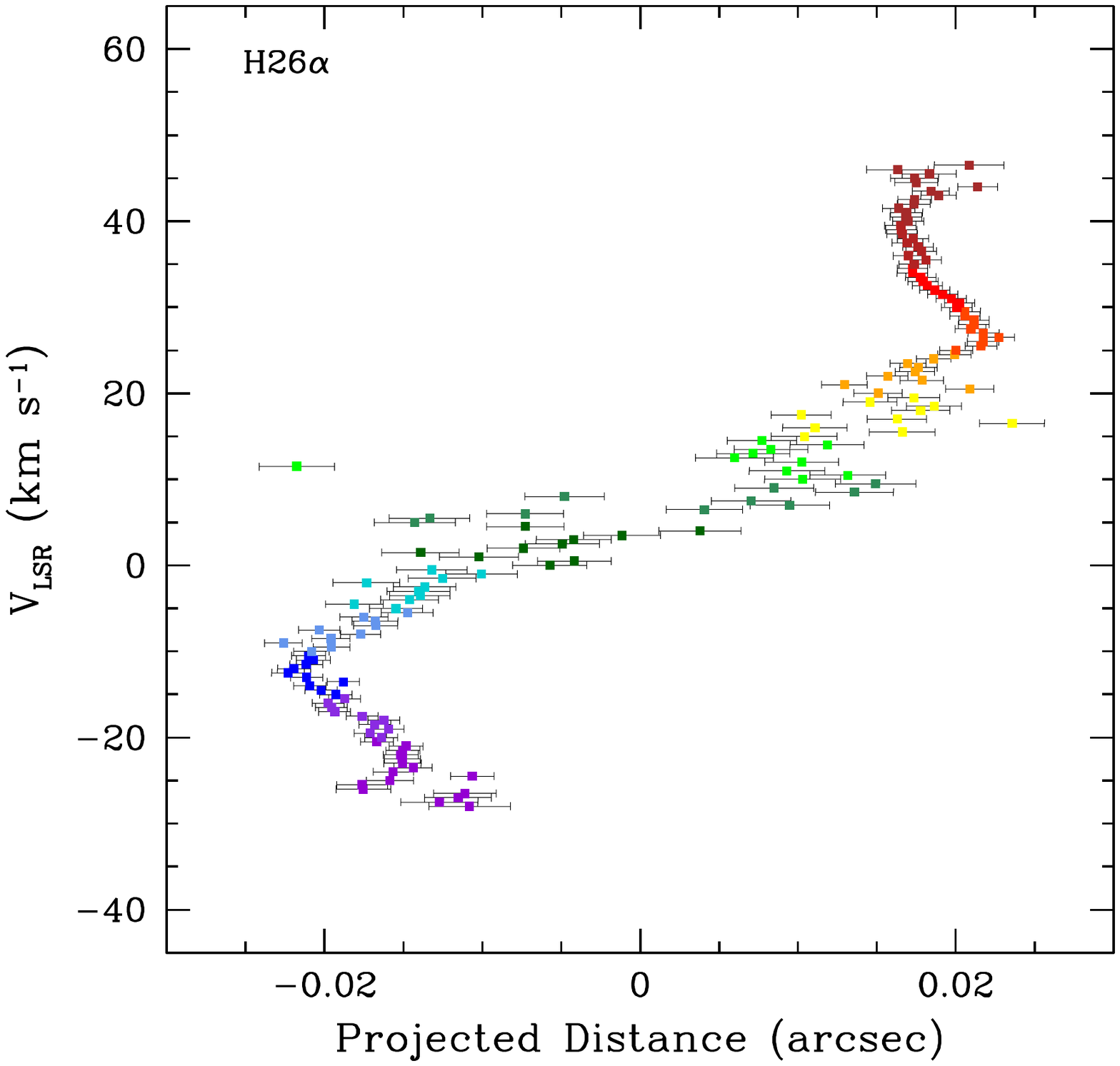} \\
\end{tabular}
\caption{\label{fig:outflow}
Rotation curves of the H30$\alpha$ ({\bf left panel}) and H26$\alpha$ ({\bf right panel}) line. The projected maser distance $l$ is computed as $\sqrt{\Delta \alpha^2 + \Delta \delta^2}$ with respect to the geometric center marked in Figure 3. The bars denote the $\pm 1\sigma$ position error from the centroid fitting and passband calibration. 
}
\end{figure}

\begin{figure}[ht!]
\includegraphics[width=15cm]{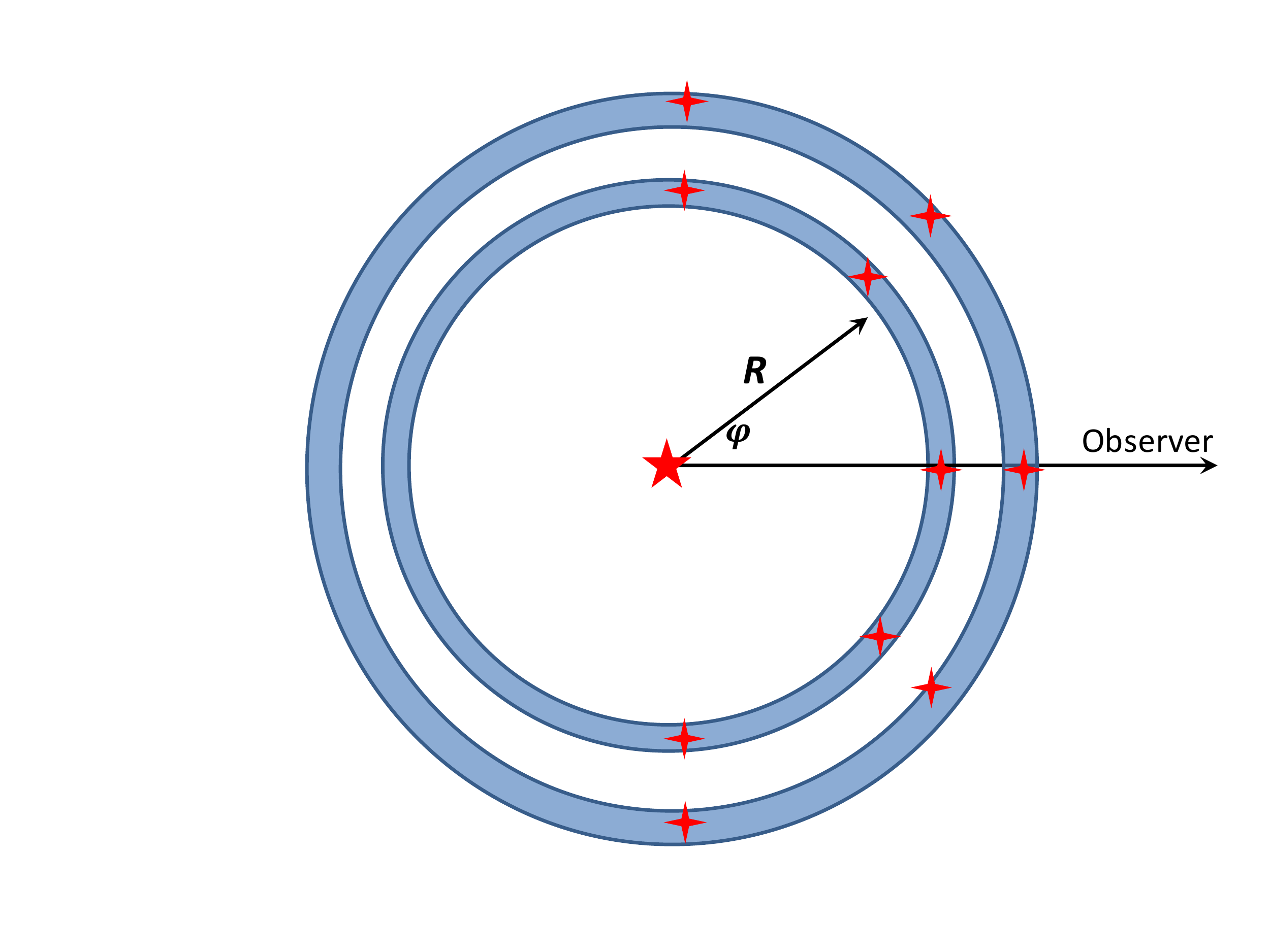} 
\caption{
A cartoon illustrating the locations of H26$\alpha$ and H30$\alpha$ masers on the disk. The red crosses represent the hydrogen recombination masers. The central star marks the position of the young stellar object. The H26$\alpha$ masers are located at the inner part of the disk of a radius of $0.021''$ (25 AU), whereas the H30$\alpha$ masers are located at the outer part of the disk of a radius of $0.025''$ (30 AU). Additional masers in the wind are not depicted here.
}
\end{figure}

\begin{figure}[ht!]
\includegraphics[width=13cm]{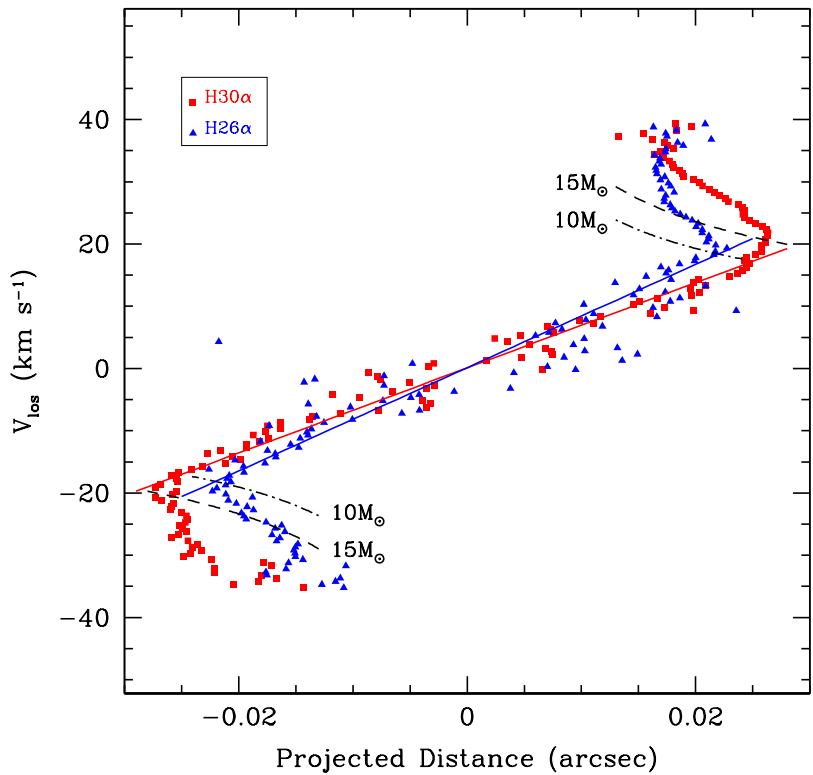} 
\caption{
Comparison of rotation curves of the H30$\alpha$ ({\bf red squares}) and H26$\alpha$ ({\bf blue triangles}) maser emission. The projected maser distance $l$ is computed as $\sqrt{\Delta \alpha^2 + \Delta \delta^2}$ with respect to the geometric center marked in Figure 3. $V_{\mathrm{los}}$ is the maser line-of-sight velocity relative to the systemic velocity of 7.2 \kms-1.
The red and blue lines represent the least-squares fits to the rotation curves of the H30$\alpha$  and H26$\alpha$ masers, respectively. The black dash-dotted lines and dashed lines represent Keplerian rotation with a central mass of 10 and 15 \msun, respectively.
}
\end{figure}


\end{document}